# Formation of carbon nano and micro structures on $C_1^+$ irradiated copper surfaces

Shoaib Ahmad

*Government College University, CASP, Church Road, Lahore 54000, Pakistan*

*National Centre for Physics, Quaid-i-Azam University Campus, Shahdara Valley, Islamabad, 44000, Pakistan*

E mail: sahmad.ncp@gmail.com

## Abstract

A series of experiments has identified mechanisms of carbon nano- and micro-structure formation at room temperature, without catalyst and in the environment of immiscible metallic surroundings. The structures include threaded nano fibres, graphitic sheets and carbon onions. Copper as substrate was used due to its immiscibility with carbon. Energetic carbon ions ($C_1^+$) of 0.2-2.0 MeV irradiated Cu targets. Cu substrates, apertures and 3 mm dia TEM Cu grids were implanted with the carbon. We observed wide range of μm-size structures formed on Cu grids and along the edges of the irradiated apertures. These are shown to be threaded nano fibers (TNF) of few μm thicknesses with lengths varying from 10 to 3000 μm. Secondary electron microscopy (SEM) identifies the μm-size structures while Confocal microscopy was used to learn about the mechanisms by which $C_1^+$ irradiated Cu provides the growth environment. Huge carbon onions of diameters ranging from hundreds of nm to μm were observed in the as-grown and annealed samples. Transformations of the nanostructures were observed under prolonged electron irradiations of SEM and TEM. One of the immediate outcomes of our present study is relevant for the efforts to fabricate carbon onions with specific properties.

## Introduction

A novel technique of producing carbon onions with unique characteristics is reported. Carbon atoms emerging out of the edges of $C_1^+$ irradiated Cu sheets are shown to form graphitic layered networks studded with carbon onions that are non-spherical, hollow and often contain smaller ones inside.

Carbon onions, the multishelled fullerenes, were discovered by intense electron beam irradiation of soot[1] with their structural manipulation reported later[2], have also been produced by $C_1^+$ implantation[3-5] and other high temperature techniques[6-8]. Their remarkable spherical shelled nanostructures found few applications; the difficulty of their structural manipulation lies in the perfect shelled symmetry. In the experiments reported here we show that energetic $C_1^+$ irradiated Cu sheets with vertical holes allow C atoms at the end of their implantation range to emerge out and form nanostructures along the edges of the holes. This technique for the formation of graphitic nanostructures that include C onions[1] and networks of graphene[9,10]. It is different from similar C onion growing methods that utilize $C_1^+$ implantations[3-8] and the graphene formation techniques that include CVD[11,12] and ion implantation[13-16] where C addition is from outside onto a surface. We do not utilize the as-deposited C ions but the atoms that emerge out of the irradiated Cu sheets' edges and sharp corners. These are shown to produce a variety of structures. This is the fundamental difference in our growth technique with that of the others. Carbon onions with various shapes and structures have been seen emerging with the under and over-lying graphene layers at all stages of C accretion and structure formation. The continuous $C_1^+$ irradiation at a steady rate is seen to produce stacks of graphene layers and nanotube-like structures. Discontinuous or variable $C_1^+$ irradiation rate may produce hollow onions that may be connected with each other or take the form that we identify as hollow onions. The hollow onions may contain smaller ones: the onion-inside-onions. Extremely low solubility of C in Cu at all temperatures and the immiscibility gap[17,18] is the key for the growth of C nanostructures on Cu surface or within. In almost all ion implantation experiments[3-5, 13-15] the $C_1^+$ dose at a given energy is the main parameter along with high substrate temperature ~700-1000 K. Their reported[3-5] critical $C_1^+$ dose for the formation of C onions ~$10^{17}$ C cm$^{-2}$. On the other hand, we have implanted ~$5\times10^{14}$ C cm$^{-2}$ at room temperature and found that the $C_1^+$ implantation rate rather than the dose is the most crucial parameter.

## Experimental Method

$C_1^+$ is implanted into Cu grids for TEM (3 mm diameter) and 120 μm thick sheets. $C_1^+$ ions are delivered by GCU's 2 MV Pelletron. The energy of ions $E(C_1^+)$ is in the range 0.2 – 0.8 MeV. We have maintained the total dose at all irradiation energies ~ $5\times10^{14}$ $C_1^+$ cm$^{-2}$. However, the rate of irradiation can be varied. The range of penetration of $C_1^+$ with these energies ~ 0.24-0.8 μm. The implanted C atoms at the end of the range, if allowed to emerge out from the sides of vertical holes in Cu sheets, are observed to form carbon nanostructures. We have exploited this peripheral aspect of $C_1^+$ irradiation whereby a fraction of C atoms escape the entrapment and emerges out of the Cu surface. These escaping C atoms are the feed of our nanostructures that are described in this letter. The carbon nanostructures so formed include multi-layered graphene stacks, nanotube-like and carbon onions. Initial investigations were done on 10 μm thick TEM Cu grids with square and circular holes. The later experiments were done on 3 mm diameter 120 μm thick Cu grids with ~ 1 mm inner holes for the nanostructure formation along the inner edges. The implantations were done at room temperature without any catalyst. Diagnostics were conducted with SEM (JEOL JSM-6480LV), Confocal microscope (Olympus FV1000) and TEM (JEOL JEM1230). One can generate a broad range of C nanostructures on 3 mm diameter, 10 μm thick Cu grids for TEM by irradiating with 0.2 MeV $C_1^+$ ions. In this paper, we present and discuss results obtained from 120 μm thick Cu rings with 3 mm OD and ~1 mm ID. Here, the initial batch of graphitic layers, sheets and shells are formed on the edges of the inner hole. These are lifted outwards by the next batch growing underneath by utilizing the C atoms emerging from within the Cu surface. The Confocal microscopic studies of the irradiated Cu grids and sheets reveal that all sharp surfaces of the holes, scratches and the edges of the grid bars are laced with C nano and microstructures. The emerging C atoms at the end of their implantation range with almost no kinetic energy are the basic ingredient and the requirement for C-C bonds to form. By exploiting this novel characteristic feature of the energetic $C_1^+$ irradiation of Cu our present technique was developed.

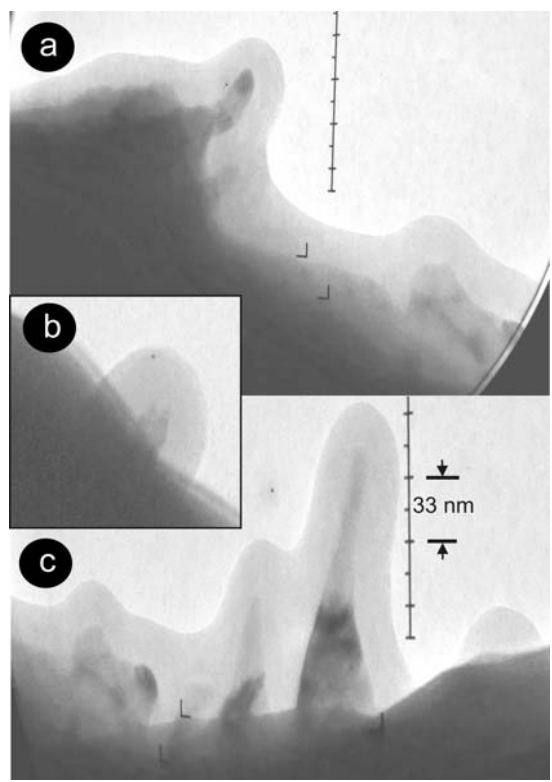

Figure 1. Earlier formed graphene layer pushed outwards by later growth. (a) Two partially spherical graphene layers emerging out of a dense background. There is a continuity of surfaces with varying heights and curvature. (b) A half spherical graphitic ball is shown that grew out of the underlying layers. (c) Four peaks of graphene layers; one on the extreme right is similar to (b); the highest multi-walled nanotube-like structure with two adjoining ones is part of the same extended structure. The scale bar's smallest unit is 16.5 nm.

## Results and Discussion

Figure 1 shows TEM micrographs of graphitic layers with variable curvature and dimensions. Figure 1(a) has spherical layered structures that can be seen superimposed on a dense layered background. There is a continuity of the outer curved surfaces of ~25 nm. Figure 1(b) shows a half spherical, graphitic ball that grew earlier than the underlying graphene layers. Four peaks of a continuum of the curved graphene layers that seem to have adjusted to the variable rate of outward flow of C atoms from below can be seen in (c). Figure 1(a)-(c) show the Cu surface decorated with multiple graphene layers. The inner conical regions of variable densities are visible within all of the outward protruding structures. The inner regions point to the routes through which C accretion occurred. Much darker layers underneath may be due to the failure of the later C addition, most probably with different rates of C outflow, to follow the earlier pattern of growth.

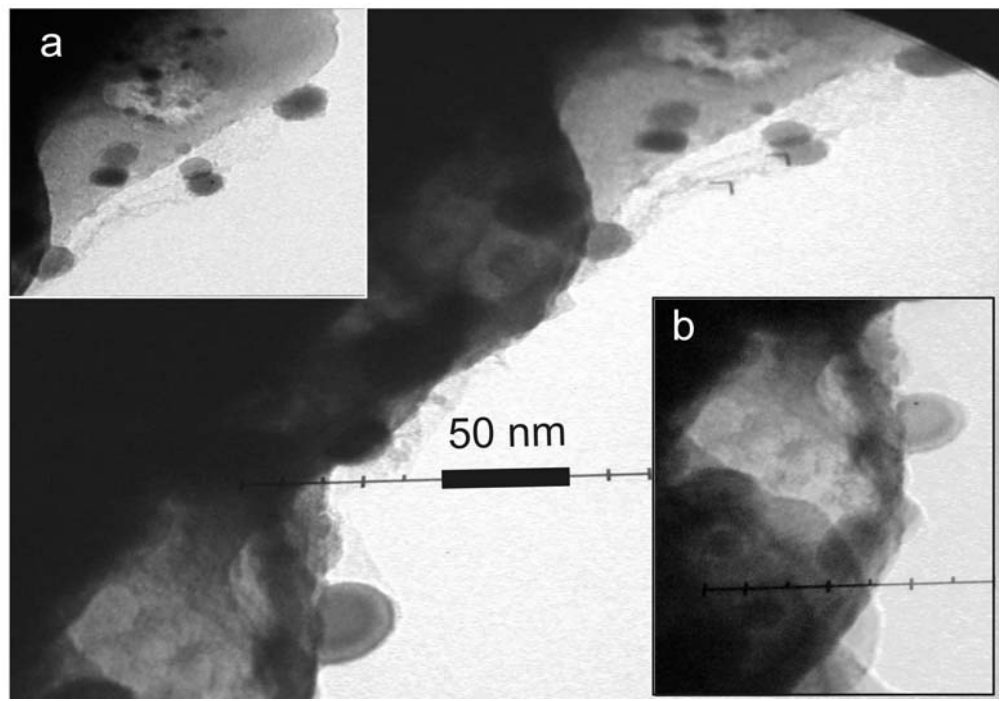

Figure 2. Hollow C onions and networks of graphitic layers. A heap of graphene layers with different orientations, thicknesses and contours is shown studded with C onions of 5-30 nm diameters. (a) The onions laced on various graphene layers in the upper right hand section are shown with sample tilt to highlight the six 10-15 nm elliptic onions and a number of smaller ones. (b) Overlapping and interconnecting mounds of graphene layers (~ 70 nm) with a protruding onion and some contained inside the mound are shown.

A graphitic stack of irregular shape with spheroidal cages of dimensions between ~5 to 30 nm having different densities and curvature can be seen in Figure 2. Most of the cages have elliptic shapes. The cages are located on various layers of graphene with perforations around the cages especially around those that protrude out of the underlying graphitic network. The closed cages are identified as C onions with many having inner structures that have different curvatures and variable inner densities. Figure 2(a) elaborates the upper right hand portion of the main figure by a sample tilt to highlight the six larger (15-20 nm) and about a dozen smaller (≤5 nm) elliptic C onions. The perforated graphene layers between the outermost onions are also visible. Figure 2(b) shows overlapping and interconnected quasi-spherical mounds of graphene layers that are laced with onions grown inside and the one that is protruding outwards. The largest, outward protruding onion shows inner structure indicating variable shell density of the onion. The diversity of the sizes and their respective positions on the graphene layers in Figure 2 can be contrasted with that shown in Figure 3(a) where a large area ~1 $\mu m^2$ is shown with various inter-connected C structures including the onions, layers and curtain-like networks of graphene sheets. The darker regions in the upper and lower halves are the dense graphitic layers studded with few quasi-spherical objects (~100 nm) and lots of smaller onions spread all over. The two boxed areas are enlarged in (b) and (c), respectively. The range of the clearly visible onions is 20-40 nm in (b) and (c) where a much larger number of smaller ones are also present. The connectivity between the distinctly shaped and separately positioned onions is presumed to be through the broken and perforated graphene layers. The diffraction spots from the centre of (c) are shown in (d) that confirms the crystallinity of the graphitic shells, however, due to the inhomogeneous onion shapes and the associated variability of the curvature, the ordering of the spots is difficult.

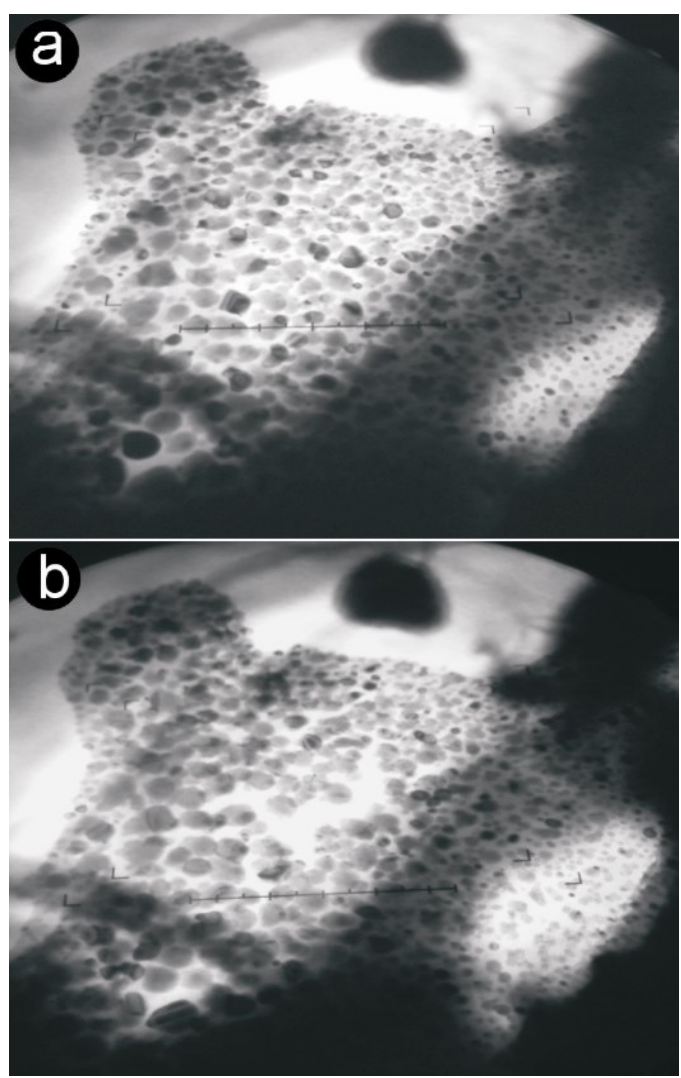

Figure 3. C onions of variety of shapes strung on multiple layers of graphene. C onions of 20-30 nm diameter with a variety of shapes are spread out on layers. (a) An area of ~ 1μm$^2$ shows a rich variety of inter-connected graphene layers, curtains and large and small C onions. Two boxed regions are further enlarged. Scale bar is 500 nm.(b) The boxed region shows a large number of C onions. (c) A large collection of ~100 C onions on relatively flatter layers with their shapes varying enormously. Scale bar is 50 nm. (d) Diffraction pattern from the centre of the onions sheet in (c) showing large number of spots indicating the presence of distinct objects with graphitic crystallinity.

We investigated with 120 keV TEM the possibility of transformations induced by intense energetic electron beam irradiations on graphitic structures with special emphasis on C onions. Intense electron beam induced changes on the arrays of onions shown in Figure 3(c). It is shown that the changes induced in onions that are spread on sheet-like graphitic networks occur via the reformation of the outer onions as well as those of the inner ones as a result of C removal. These changes were observed to occur continuously during the electron beam irradiations with many C onions retaining their identity as individual onions while a few did merge into each other. On the other hand, the effect of C removal and the associated structural readjustments appeared as significant in C onions that are connected with each other as depicted graphically in Figure 4 (a) and (b).

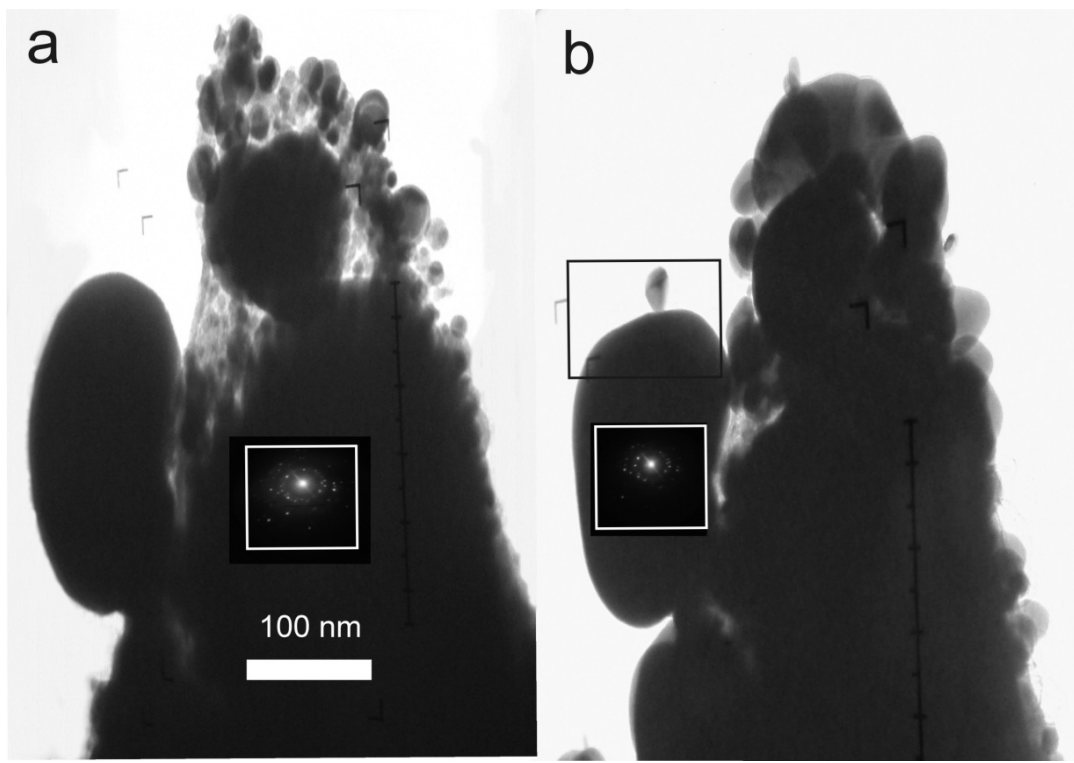

Figure 4. Coalescence of smaller C onions into larger ones and the shape changes of the larger graphitic cages. (a) A collection of three large (≥50 nm) and many smaller C onions, all sticking together in their pristine positions is shown in the first TEM micrograph. (b) After intense 120 keV electron beam irradiation in TEM for 1 minute the smaller ones coalesce into larger C onions that have relatively well defined surfaces. White boxes show the diffraction spots from the center of the graphitic caged regions. The scale bar is 100 nm.

The transformations induced in large (≥100 nm) and small (≤20-40 nm) cages is shown to occur via coalescence of the cages as is clear by comparing 4(a) with 4(b). In 4(a) two large cages and many smaller elliptic ones are connected with a much larger (~200 nm diameter x 500 nm length) graphitic caged structure. Under intense 120 keV electron beams the smaller onions on top of the structure in 4(a) coalesced into the larger ones shown in 4(b). Coalescence of smaller onions has produced comparatively larger spheroidal structures with increased densities. An observable feature of the reformed C onions is that these remain connected with each other. Major changes also occurred in a large kidney shaped cage that can be seen on the left in 4(a) prior to being subjected to intense 120 keV electron beam. The noticeable changes seen are in the outward convex curvature supplemented by the appearance of two types of nanostructures that are enlarged in Figure 5(a). The two largest structures in Figure 4 show diffraction spots from the beam that was focused at the point where the pattern is embossed. One can see that crystallinity of the objects is graphitic with wide range of overlapping planes in both the cages that seem to contain many smaller spheroidal, onion-like structures.

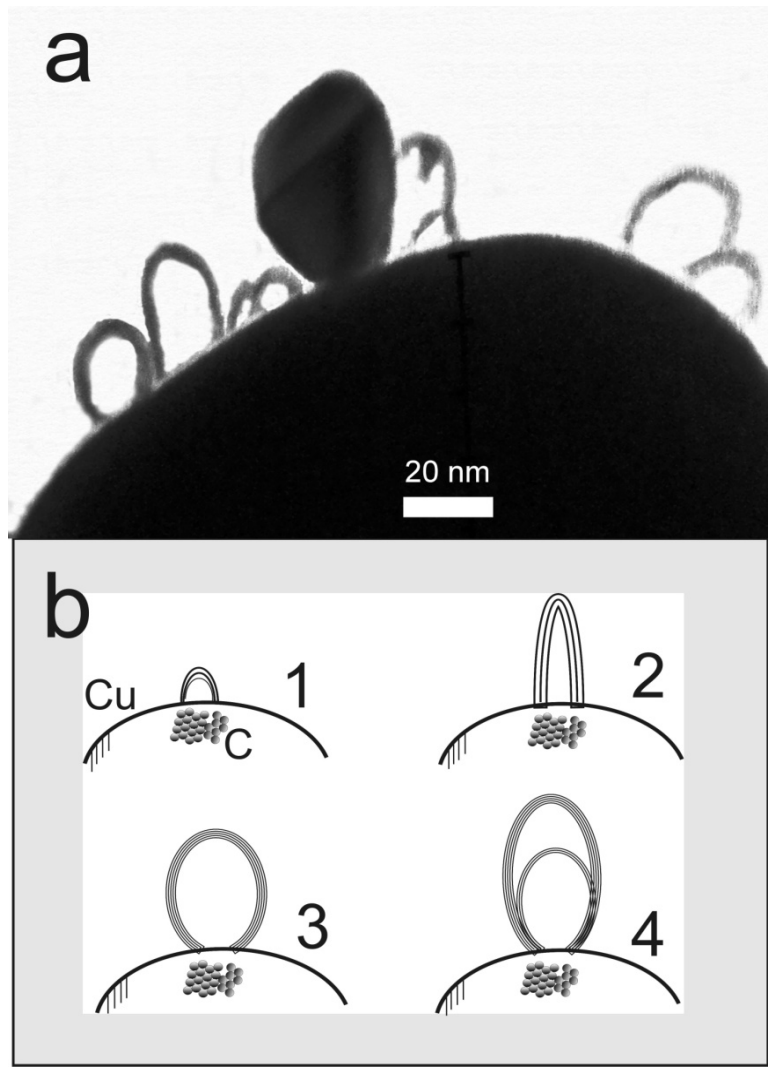

Figure 5. In situ formation of onion-inside-onions and nanotubes. (a) The boxed region in Figure 4(b) is further enlarged to reveal six new, multi-walled nanotubes of ~2-3 nm thick shells. These emerged out of the reshaped, larger graphitic caged structure under energetic electron irradiations. (b) Four proposed stages (1-4) of the formation of multi-walled nanotubes or multi-shelled onion-inside-onions are graphically represented. The emergence of C from under the Cu surface defines this mechanism of formation that is inward growth as opposed to the usual outward one.

In Figure 5 the boxed portion of Figure 4(b) is enlarged where the central ellipsoidal that is a C onion-inside-onion in the making with ~30 nm x 45 nm dimensions. Six new hollow multiwalled nanotube-like objects with 5-7 shell thickness and different heights can be seen. This in situ observation of the different stages of C onion and nanotube growth out of the graphitic surface leads us to similar mechanisms of formation by the C atoms emerging from within the irradiated Cu surface. The nanotube-like features, onions and multiple layered stacks of graphene shown in Figures 1 and 2 can be understood by the comparing with the energetic electron induced transformations in Figure 5(a). In 5(a) two distinct but structurally related C nanostructures are formed by the C atoms that are released by 120 keV electrons from the underlying graphitic surface. The graphical representation of the growth mechanism is presented in four sequences 1-4 in Figure 5(b). The structure underneath the irradiated region and the rate of flow of the emerging C atoms determine the curvature and the lateral extent of the newly forming C shells (step 1). For a steady flow of C, shells-inside-shells grow without significant variations in diameter resulting in nanotube-like features (step 2). For increasing outward flow of C one is likely to obtain onion-like structure (step 3). Discontinuous or variable C flow rates are likely to yield onion-inside-onions (step 4). Majority of the C onions seen in Figure 3 belong to this category. Such an object can also be seen in the TEM micrograph 5(a). This in situ growth of an onion-inside-onion and the six hollow nanotubes is due to the C atoms released by the 120 keV electrons from the graphitic surface of the C cage. One finds similarities by comparing the C onions formed by C atoms that emerge from the irradiated Cu surfaces or the ones that are released by breaking the $sp^2$ bonds of the graphitic structures by 120 keV electrons.

## Conclusion

We show for the first time, hollow, onion-inside-onion structures. These are formed by the C atoms emerging from $C_1^+$ irradiated Cu surface at room temperature. The rate of $C_1^+$ irradiation and the route that C atoms take to escape from within the Cu lattice determines the formation of C onions and graphene layers. Spherical and cylindrical curvature as discussed elsewhere[19] is seen as the essential feature by which the C onions and graphene layers accommodate variations in the C accretion rate and the Cu surface irregularities at the region of C emergence. Pentagons and heptagons may be forming along with the hexagonal networks to yield the diversity of curvature in the layers and shells of onions. Formation at room temperatures forbids the Stone Wales[20] transformation to smooth out the sharp corners by bond rearrangements.

In conclusion we have shown that the $sp^2$-bonded nanostructures formed by C atoms emerging from within, onto an immiscible surface like Cu, will include all possible configurations including multiple layered graphene, nanotubes and onions. Their respective number densities and shapes, however, depend on the rate of emergence which is linked with the ion irradiation rate and of course, the local surface configurations at the point of C atoms' emergence. We have proposed a growth mechanism by demonstrating in situ growth of C cages. Carbon onions with well defined shell thickness and inner radii have been shown to be a possibility and the onion-inside-onion configuration can be exploited to yield desired physical properties.

### REFERENCES

1. Ugarte, D. Curling and closure of graphitic networks under electron-beam irradiation. *Nature* **359**, 707-709 (1992).
2. Banhart, F. & Ajayan, P. M. Carbon onions as pressure cells for diamond formation. *Nature* **382**, 433-435 (1996).
3. Cabioch, T. Riviere, J. P. & Delfond, J. a new technique for fullerene onion formation. *J. Mater. Sci.* **30**, 4787-4792 (1995).

4. Cabioch, T. *et al.* Carbon onion formation by high-dose carbon ion implantation into copper and silver. *Surf. Coatings Technol.* **128-129**, 43-50 (2000).
5. Abe, H. Nucleation of carbon onions and nanocapsules under ion implantation at high temperature. *Diamond and related Mater.* **10**, 1201-1204 (2001).
6. Sono, N. Wang, H. Chhowalla, M. Alexandou, I. & Amaratunga, G. A. Synthesis of carbon onions in water. *Nature* **414**, 506-507 (2001).
7. Szerencsi, M. & Radnoczi, G. The mechanism of growth and decay of carbon nano-onions formed by ordering of amorphous particles. *Vacuum* **84**, 197-201 (2010).
8. Blank, V. D. Kulnitskiy, B. A. & Perezhogin, I. A. Structural peculiarities of carbon onions, formed by four different methods: onions and diamonds, alternative products of graphite high pressure treatment. *Scripta Materilia* **60**, 407-410 (2009).
9. Meyer, J. C. Geim, A. K. Katsnelson, M. I. Novosolov, K. S. Booth, T. J. & Roth, S. The structure of suspended graphene sheets. *Nature* **446**, 60-67 (2007).
10. Geim, A. K. Graphene: Status and Prospects. *Science* **324**, 1530-1534 (2009).
11. Mathevi, C. Kim, H. & Chhowalla, M. A review of chemical vapor deposition of graphene on copper. *J. Mater. Chem.* **21**, 3324-3334 (2011).
12. Reina, A. *et al.* Large area, few-layered graphene films on arbitrary substrates by chemical vapr deposition. *Nano. Lett.* **9**, 30-35 (2009).
13. Bangert, U. Bleloch, A. Gass, M. H. Seepujak, A. & van der Berg, J. Doping of few-layered graphene and carbon nanotubes using ion implantation. *Phys. Rev. B* **81**, 245423-245434 (2010).
14. Garaj, S. Hubbard, W. & Golovchenko, J. A. Graphene synthesis by ion implantation. *Appl. Phys. Lett.* **97**, 183103-183107 (2010).
15. Baraton, L. *et al.* Synthesis of few-layered graphene by ion implantation of carbon in nickel thin films. *Nanotechnology* **2**2, 085601-085605 (2011).
16. Sun, Z. *et al.* Growth of graphene from solid carbon sources. *Nature* **468**, 549-552 (2010).
17. Fuks, D. *et al.* Carbon in copper and silver: diffusion and mechanical properties. *J. Mol. Structure (Theochem)* **539**, 199-214 (2001).
18. Sun, J. & Zhang, M. D. Interface characteristics and mechanical properties of carbon fibre reinforced copper composites. *J. Mater.Sci.* **26**, 5762-5766 (1991).
19. Ahmad, S. Criteria for the growth of fullerenes and single-walled carbon nanotubes in sooting environments. *Nanotechnology* **16**, 1739-1745 (2005).
20. Stone, A. J. & Wales, D. J. Theoretical- studies of icosahedral $C_{60}$ and some related species. *Chem. Phys. Lett.* **128**, 501-503 (1986).

## ACKNOWLEDGMENTS

Authors acknowledge financial support by the Higher Education Commission of Pakistan (HEC) to set up the 2 MV Pelletron and electron microscopy labs at Government College University (GCU), Lahore. We acknowledge Mr. M. Khaleel for technical support with the Pelletron and Ms. Farwa Nurjis of NIBGE, Faisalabad for Confocal Microscopy.